# Surface and bulk properties of the Heusler compound $Co_2Cr_{0.6}Fe_{0.4}Al$: a Mössbauer study

**Vadim Ksenofontov, Sabine Wurmehl, Sergey Reiman, Gerhard Jakob and Claudia Felser**

Institut für Anorganische Chemie und Analytische Chemie, Johannes Gutenberg-Universität, Staudinger Weg 9, D-55099 Mainz, Germany

E-mail: v.ksenofontov@uni-mainz.de

**Abstract.** To explore its structural and magnetic properties, the Heusler compound $Co_2Cr_{0.6}Fe_{0.4}Al$ was investigated using Mössbauer spectroscopy. The results of both transmission and conversion electron Mössbauer spectroscopy (CEMS) are analyzed to obtain insight into both the disorder effects as well as the differences between bulk and surface properties. It was found that mechanical treatment of the surfaces of bulk samples causes disorder and phase segregation, effects that should be taken into consideration when performing studies using surface-informative techniques. Results from bulk sample CEMS measurements of $Co_2Cr_{0.6}Fe_{0.4}Al$ are used to interpret the thin film Mössbauer spectra of this compound.

PACS numbers: 75.50.Cc, 72.25.Ba, 71.20.-b, 82.80.Ej

*Keywords*: Heusler compounds, Mössbauer spectroscopy

## 1. Introduction

In spintronics applications, the injection of polarized spin currents requires materials whose conduction electrons have a high spin polarization. Promising candidates for these materials are the half-metallic ferromagnets, which are characterized by the presence of a band gap in only one spin direction and by metallic properties in the other spin direction. Many Heusler compounds with the stoichiometric composition $X_2YZ$ (where $X$ and $Y$ are transition metals, and $Z$ denotes an $sp$ element) and ordered in the $L2_1$ structure are predicted to be half-metallic ferromagnets. The Heusler compound $Co_2Cr_{1-x}Fe_xAl$ is a prospective compound for achieving half-metallicity at room temperature (RT) because the electronic structure can be controlled by varying the Fe concentration. In particular, $Co_2Cr_{0.6}Fe_{0.4}Al$ (CCFA) recently aroused special interest because a relatively high negative magneto-resistance effect of up to 30 % was found in powder samples that were in a low magnetic field of 0.1 T [1-3]. The parental non-doped compound $Co_2CrAl$ is expected to exhibit a high spin polarization [1, 2, 4-8]. In experiments, however, $Co_2CrAl$ displays only one half of the magnetic moment value expected from the Slater-Pauling rule for half-metallic compounds [9]. To explain this result, Miura *et al.* proposed that the Co sites were occupied by Cr ($DO_3$ type disorder) [5, 6]. This type of disorder leads to an antiferromagnetic coupling of the antisite Cr with the nearest neighbor ordinary site Cr atoms, and strongly decreases the spin polarization value. It has been concluded that $Co_2FeAl$ bulk samples order in the $L2_1$ structure, but they can also demonstrate $B2$ with $Y$-$Z$ and $A2$ phases with $X$-$Y$-$Z$ elements occupying their sites at random [10]. Thin CCFA films have been prepared by several groups [11–14]. Hirohata *et al.* have grown $L2_1$ polycrystalline $Co_2CrAl$ and epitaxial $L2_1$-structured $Co_2FeAl$ films onto GaAs(001) substrates [12, 13]. However, according to published results, all thin CCFA films always exhibited a $B2$ structure. Using Andreev reflections, a spin polarization of approximately 49% was found for polycrystalline CCFA samples [15]. Clifford *et al.* recently reported a spin polarization of 81% in CCFA point contacts [16]. The observed incomplete spin polarization may be caused not only by atomic disorder but also by the surface properties of the samples studied in the spin polarization experiments. At RT, CCFA sputtered films with the predominantly $B2$ structure exhibit a tunnel magnetoresistance (TMR) of 19%, which corresponds to a spin polarization of 29% [17]. This decrease in spin polarization from that expected for a half-metallic compound may be due to the presence of atomically disordered phases, the non-stoichiometric composition of the sputtered films, or the rough interfaces between the CCFA and the Al–O tunnel barrier. Fully epitaxial $CCFA/MgO/Co_{50}Fe_{50}$ magnetic tunnel junctions that have a nearly stoichiometric composition exhibit TMR ratios of 90% at room temperature and 240% at 4.2 K [18].

XMCD studies have indicated that the surface properties are different from the properties of bulk compounds [19]. These studies indicated the small magnetic moment of Cr atoms and the equal coordination of Fe atoms in powder and bulk samples. In contrast, Co and Cr atoms demonstrated different local surroundings in powder and bulk probes. It should be mentioned that before the XMCD measurements were made, the surfaces were scratched with a diamond file *in situ* while under ultrahigh vacuum to remove the surface oxide layer.

It appears that the surface states of particles must be taken into account when interpreting the magnetoresistance effects in powder compacts and TMR devices. It is therefore important to clarify the phase composition of the surfaces of the particles. Routine X-ray powder diffraction measurements can not clarify the phase composition of surface layers that are less than 100 nm thick or those that display amorphization. These limitations do not exist for conversion electron Mössbauer spectroscopy (CEMS), which was the main tool in this study for examining the surface state of CCFA.

**2. Experimental section**

*2.1 Synthesis and sample preparation*

Following procedures described elsewhere, CCFA was prepared by arc melting under a purified argon atmosphere [20]. Melted samples were quenched down to room temperature by rapidly shutting off the heating. To magnify the Mössbauer signal, the iron constituent in several probes was enriched to 10% by using $^{57}Fe$. The structure of each sample was characterized using X-ray powder diffraction. Using a spark erosion machine, the ingots were cut into 1 mm thick discs. Remaining parts of the ingots were crushed and powdered to a grain size of approximately 100 μm for use in making transmission Mössbauer spectroscopic measurements.

To remove oxidized layers and imperfections on the surfaces of the sample discs, a multi-stage polishing procedure with fine grades of SiC paper, SiC powder and diamond paste was used. The mirror-like quality of the surface was achieved after 8 steps of polishing; the final polishing was performed with a diamond paste having a grain size of 1μm. After Mössbauer measurements, a 30 nm thin film of CCFA containing natural iron was deposited on the enriched CCFA sample substrate by using magnetron sputtering. The sample was measured again and a second 30 nm thick non-enriched layer was deposited on the first layer; the enriched substrate that was now covered with a non-enriched 60 nm layer of CCFA was then measured again. Finally, for the Mössbauer studies, the top layers of the bulk samples were removed by sputtering the surfaces to a depth of up 0.5 μm by using $Ar^+$ ion bombardment with an energy of 4-5 keV and a target current of 25 mA, The $Co_2Cr_{0.6}Fe_{0.4}Al$ film sample that was 100 nm thick was prepared by dc-sputtering from a stoichiometric target in a High Vacuum sputtering system with a rest pressure of $5·10^{-8}$ mbar. The $Al_2O_3$ substrate was heated to 100 °C during deposition in a high purity Ar atmosphere of $5·10^{-2}$ mbar.

*2.2 Magnetic characterization*

The magnetic characterization of the samples was performed using a Quantum Design MPMS-XL SQUID magnetometer equipped with a high temperature furnace. The Curie temperature of 760 K was determined by heating the $Co_2Cr_{0.6}Fe_{0.4}Al$ sample in an external field of 1 kOe. In this study, we found a saturation magnetization of 3.2(1) $\mu_B$/f.u.

*2.3 Mössbauer spectroscopy*

$^{57}$Fe Mössbauer resonance absorption is accompanied by the emission of gamma quanta (14.4 keV), conversion electrons (7.3 keV), Auger electrons (5.6 keV), and X-rays (6.3 keV). CEMS monitors a depth of approximately 100 nm, whereas the characteristic 6.3 keV X-ray radiation comes from a depth of approximately 1000 nm and provides information about the bulk properties. The CEMS was performed by using a He + 4% $CH_4$ gas flow CEMS proportion counter operating down to 93 K. X-ray Mössbauer spectra were measured in the reflection mode. The Recoil 1.02 Mössbauer Analysis Software was used to fit the experimental spectra [21]. Isomer shift values are quoted relative to α-Fe at 293 K.

**3. Experimental results and discussion**

Mössbauer spectroscopy was performed to clarify the magnetic state of Fe atoms in CCFA, to explain its magneto-crystalline structure, and to compare its bulk and surface properties. Figure 1(a) shows the X-ray reflection Mössbauer spectrum of CCFA at room temperature. A hyperfine field distribution model was used for interpreting the spectrum [22]. The extracted hyperfine magnetic field distribution (Figure 1(b)) shows a main asymmetrical broad peak at a hyperfine field of 291 kOe, and two weak peaks at 172 kOe and 3 kOe. According to the results from transmission $^{57}$Fe Mössbauer spectroscopic studies of powdered samples reported by Wurmehl et al. [23], the spectrum with a hyperfine magnetic field of approximately 290 kOe has to be considered as being the main fingerprint for iron atoms in bulk ordered CCFA.

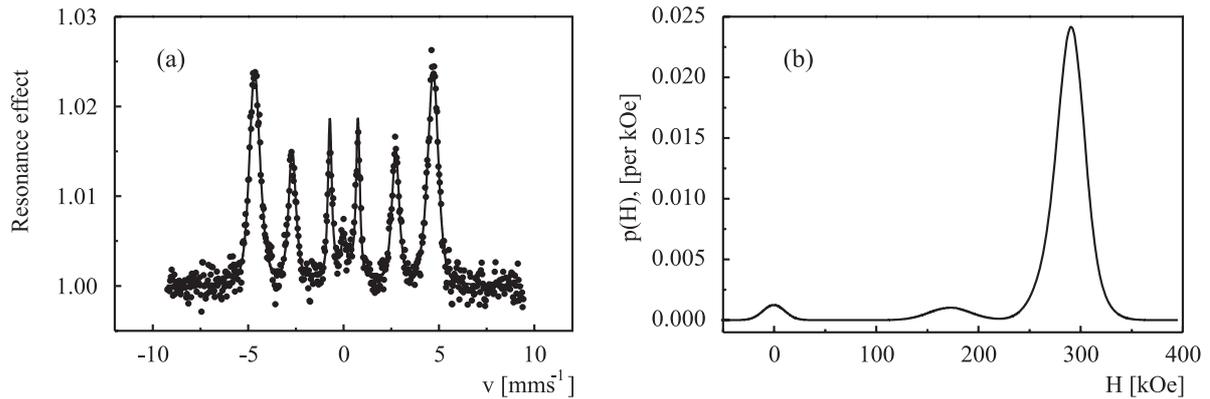

Figure 1. X-ray reflection Mössbauer spectrum (a) and the extracted hyperfine field distribution (b) in bulk CCFA.

CEMS results for bulk CCFA samples at T = 293 K and T = 95 K that have been roughly polished with sand paper are presented in Figure 2(a, b). The component in the hyperfine distribution (T = 293 K) at an $H_{hf} \cong 15$ kOe attains almost 25% of the total spectral intensity. The ratio of the intensity of this line with the next two peaks located at 89 kOe and 236 kOe is approximately 1:1:2. It appears that the part of the distribution centered at $H_{hf} \cong 286$ kOe stems from the 12% fraction of iron atoms in the bulk CCFA. The additional signals that appear in the CEMS spectrum of the roughly polished sample are evidence of different iron positions that may be due to a phase separation caused by the polishing procedure. In temperature dependent CEMS measurements at 95 K, the fraction of bulk CCFA is conserved, whereas the ratio of subspectra intensities for $H_{hf} \cong 17$ kOe, $H_{hf} \cong 99$ kOe and $H_{hf} \cong 250$ kOe becomes 0.6:1:2. The low temperature CEMS measurements and the CEMS measurements in an external magnetic field of 1.3 T did not display any superparamagnetic behavior of the surface layers that were affected by mechanical treatment. It can be concluded that rough mechanical treatment (polishing, scratching) of the surface causes a phase separation or segregation of the components that comprise CCFA. Since the formula for the bulk compound is given by $Co_2Cr_{0.6}Fe_{0.4}Al$, the subspectra with hyperfine fields lower than 20 kOe can be attributed to a Cr-reach phase, while the subspectra with an average $H_{hf} \cong 100$ kOe arise from an Al-reach phase. The average value $H_{hf} \cong 250$ kOe of a Co-reach phase is slightly less than the hyperfine field of bulk CCFA. The contribution of the Fe-reach phase is not seen in Mössbauer spectra since it arises from the minority component.

The site-specific magnetic moments obtained by XMCD measurements for *in situ* scratched samples [19, 24, 25] can be explained as being caused by the formation of an almost non-magnetic Cr-reach phase from the rough polishing. Indeed, experimentally found magnetic moments at the Cr site are too low when compared to calculations.. At the same time, the XMCD measurements show a moderate decrease of the magnetic moments associated with Co atoms, a result that is in agreement with the attribution of $H_{hf} \cong 250$ kOe to a Co-reach phase as discussed above.

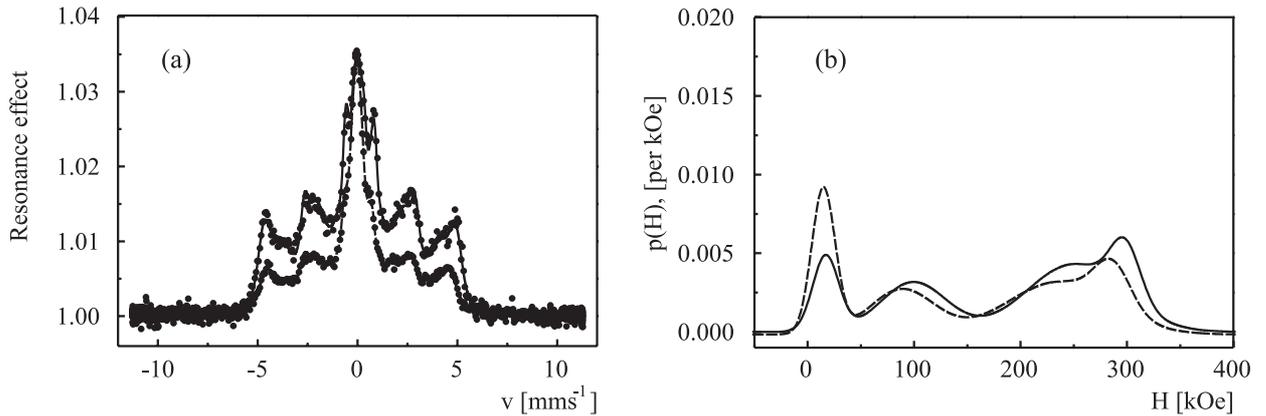

Figure 2. CEMS spectra of bulk polished CCFA (a) and the extracted hyperfine field distribution (b) at RT (dashed line) and at 93 K (solid line).

To clarify the nature of the surface phase separation, the roughly polished surface of a bulk CCFA sample was covered twice with non-enriched CCFA by using RF magnetron sputtering. After first covering the surface with a 30 nm thick layer, the CEMS spectrum is not visibly changed from the original spectrum (Figure 3(a)). A radical difference appears after the second 30 nm thick layer was applied. In contrast to the spectra obtained after rough polishing, the CEMS spectrum shown in Figure 3(b) is dominated by a magnetic sextet and is similar to the bulk CCFA spectrum (Figure 1). The distribution of the hyperfine field plot exhibits a peak at 290 kOe, corresponding to bulk CCFA, and a peak at 13 kOe that is attributed to the Cr-reach phase. The middle part containing the Al- and Co-reach phase segregations is smeared out. It must be emphasized that the 60 nm layer is practically "invisible" in this experiment because of the non-enriched composition of the covering CCFA, and that the CEMS spectrum only reflects the state of the $^{57}$Fe enriched interface. The restoration of magnetic order while the sample surface was covered could be induced by a magnetic state of the 60 nm covering and requires future detailed experimental study. The possibility that the structural order was restored on the sample surface can be excluded because, according to the preparation procedure, CCFA is a metastable compound. Indeed, annealing CCFA leads to a degradation of magnetic properties, a degradation that is attributed to increased atomic disorder. Previous investigations showed a drastic increase in the Mössbauer transmission spectrum of a paramagnetic component after the CCFA was annealed [1]. In the cited study, the paramagnetic component was interpreted as being caused by Fe atoms occupying tetrahedral Co sites, whereas Fe occupies only octahedral sites in quenched samples.

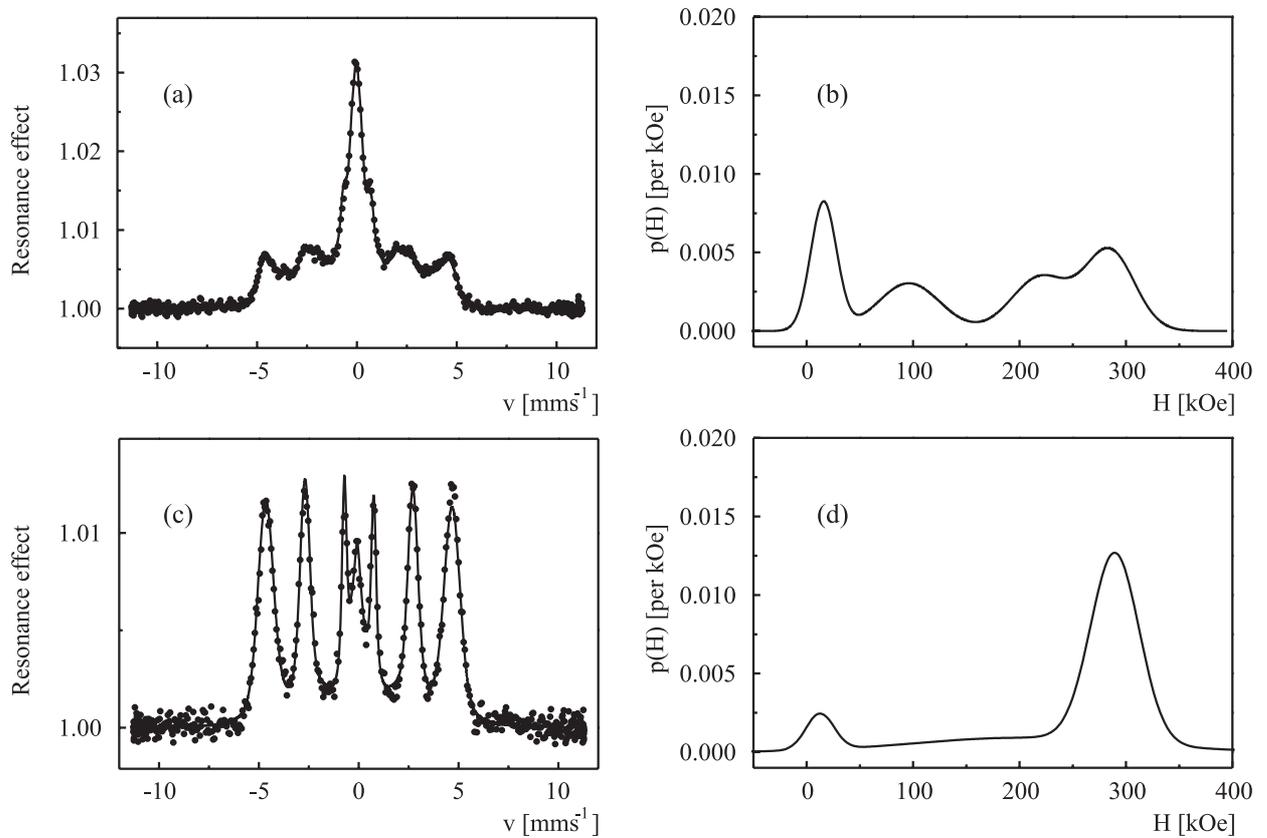

Figure 3. CEMS spectra and HF distribution of $^{57}$Fe enriched CCFA bulk material after covering with a 30 nm non-enriched CCFA thin film (a, b), and a 60 nm non-enriched CCFA thin film (c, d).

Subsequent multi-step fine machine polishing of the rough surface using diamond pastes having grain sizes from 6 μm to 1 μm removes associated with phase separation components in favor of the CCFA phase (Figure 4). The best cleansing and removal of extrinsic species from the surface was finally achieved by using Ar$^+$ ion bombardment for sputtering the top layers to a depth of up to 0.5 μm. The CEMS spectrum and HF distribution presented in Figure 5 are similar to what is found for finely polished samples, but show a higher fraction of the basic CCFA component. It should be noted that the Mössbauer measurements did not display any evidence of iron oxides although this was expected after transportation of the cleaned samples from the vacuum chamber to the Mössbauer spectrometer.

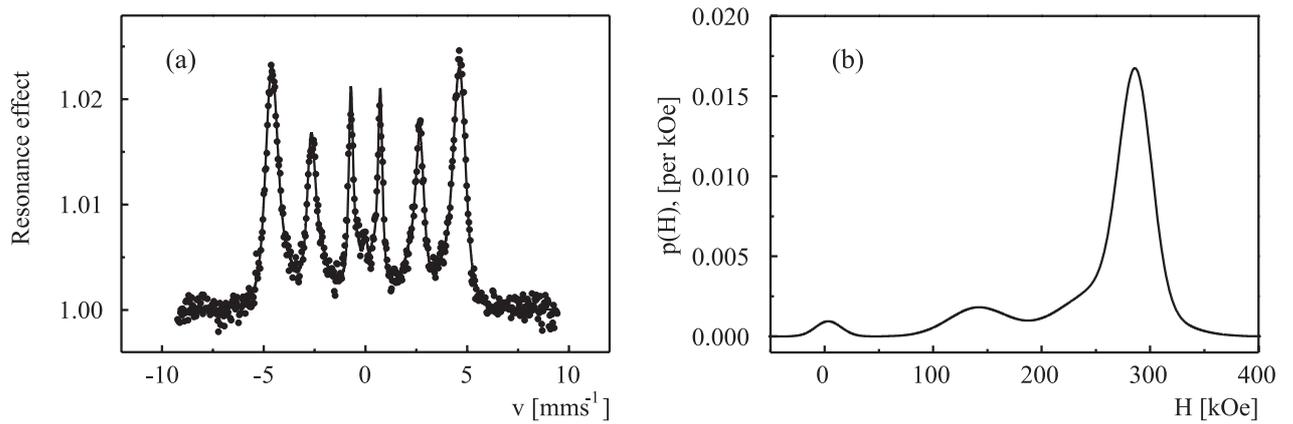

Figure 4. CEMS spectrum and HF distribution of $^{57}$Fe enriched CCFA bulk material after fine multi-step polishing.

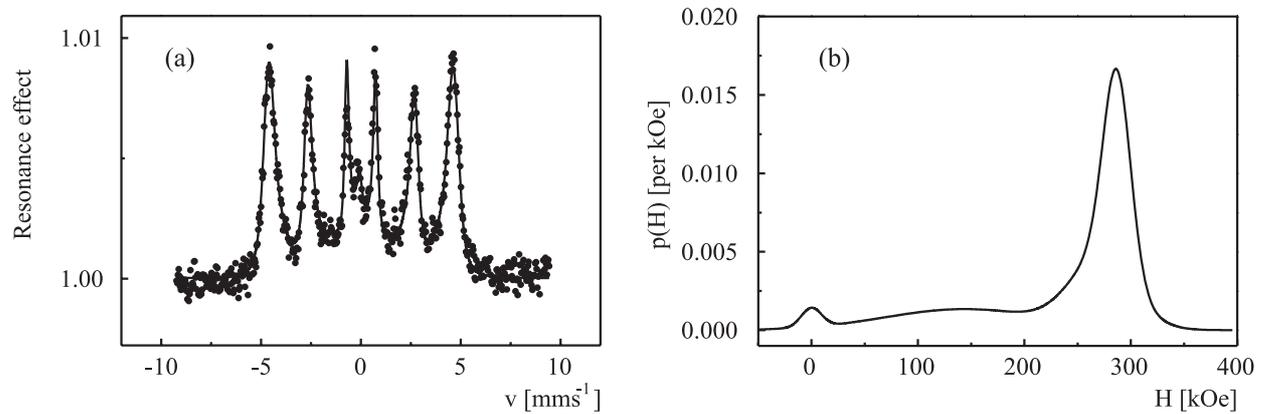

Figure 5. CEMS spectrum and HF distribution of $^{57}$Fe enriched CCFA bulk material after Ar$^+$ ion bombardment.

A detailed CEMS study of bulk CCFA samples enables us to interpret the distribution of hyperfine magnetic fields on iron atoms in thin films based on CCFA. Figure 6 shows the CEMS spectrum of a 100 nm thick CCFA film deposited on Al$_2$O$_3$ under the conditions noted above. The low value of 0.2% for the resonance effect, compared to the resonance effect in bulk CCFA, is due to the natural iron content in the thin film. The extracted HF distribution is smeared and shows two maxima centered at approximately 50 kOe and 240 kOe. Although they are not direct analogues of the surface spectra of bulk CCFA, it may be assumed that the former peak is a "fusion" of peaks characteristic of the Cr- and Al-reach phases, whereas the latter peak resembles a peak that is specific to a Co-reach phase but that is shifted down to lower fields. The simplification of the HF distribution, the broadening of the peaks and their "fusion" may be interpreted as being due to the randomization of a structure towards an *A*2 type of disorder. It appears that the expected HF distribution in thin films with an *L*2$_1$ or *B2* structure should be similar to distributions found after fine multi-step polishing (Figure 4) or Ar$^+$ ion bombardment (Figure 5).

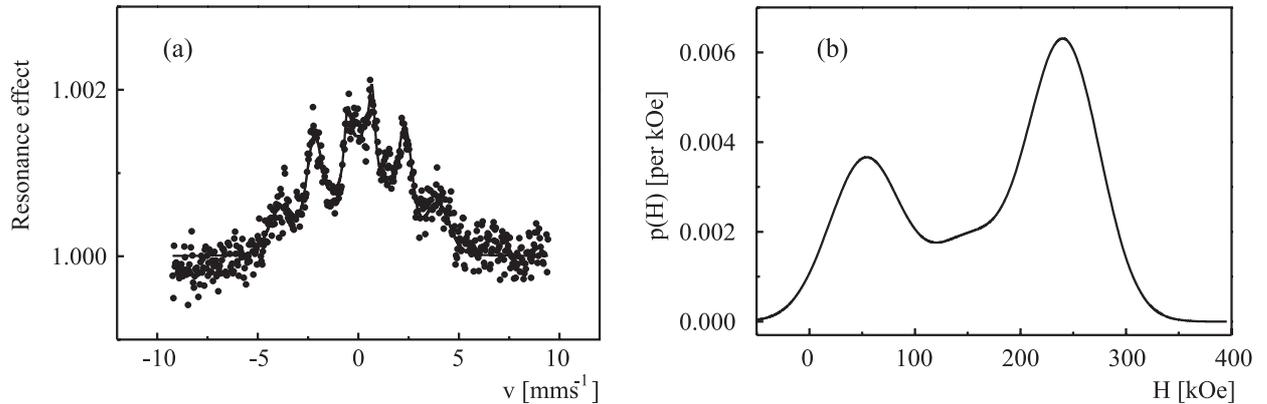

Figure 6. CEMS spectra and HF distribution of a 100 nm CCFA film deposited on $Al_2O_3$.

## 6. Conclusions

Disorder effects in the prospective half-metallic ferromagnetic compound CCFA were investigated using transmission and conversion electron Mössbauer spectroscopy (CEMS). It was found that mechanical treatment leads to a decomposition of the surface layer and a segregation of the composition in species with distinct values of hyperfine fields on the iron atoms. Disordering thus degrades the magnetic properties and causes the spin polarization in CCFA to vanish. A possible reason for the decomposition may be the metastable character of this compound. The inducing of disorder by mechanical treatment on the surfaces of bulk samples should be taken into consideration when studies using surface-informative techniques are performed; i.e. scratching and rough polishing distorts the surface of the CCFA and may potentially lead to the misinterpretation of surface phenomena. On the other hand, the presence of weakly magnetic or non-magnetic layers on the surfaces of particles should play an important role in the magnetoresistance and transport properties that are utilized in TMR devices based on CCFA. The CEMS study of bulk samples that is presented here elucidates the peculiarities that may be observed in the Mössbauer spectra of ordered thin films based on CCFA.

## Acknowledgments

We are grateful for financial support from the DFG (FG 559, TP1, 2, 8). The Materialwissenschaftliches Forschungszentrum der Universität Mainz is also gratefully acknowledged.